\documentclass[12pt,preprint]{article}
\usepackage{graphicx}
\usepackage{txfonts} 
\graphicspath{{fig/}}

\newcommand\G{\mathcal{G}}
\newcommand\V{\mathcal{V}}
\newcommand\E{\mathcal{E}}

\begin{document}
%
\title{Network Inference from TraceRoute Measurements:
        Internet Topology `Species'}

\author{Fabien~Viger$^1$,
        Alain~Barrat$^2$,
        Luca~Dall'Asta$^2$,\\
        Cun-Hui~Zhang$^3$,
        and Eric~D.~Kolaczyk$^{4,}$\thanks{Contact Author:
        Eric Kolaczyk ({\it kolaczyk@math.bu.edu}), Department of
        Mathematics and Statistics, Boston University, Boston, MA
        02215, USA.}
        }

\markboth{IEEE Journal on Selected Areas in Communications\\ Sampling
the Internet: Techniques and Applications, 2006}{Inferring the size of
the Internet from partial traceroute measurements}

\maketitle

\begin{center}
\small{$^1$LIAFA, CNRS and Universit\'e de Paris-7, and LIP6,
        CNRS and Universit\'e de Paris-6, France\\ 
$^2$ LPT and UMR 8627 du CNRS, Universit\'e de Paris-Sud, France\\
$^3$ Department of Statistics, Rutgers University, USA \\
$^4$ Department of Mathematics and Statistics, Boston University, USA}
\end{center}

\begin{abstract}
Internet mapping projects generally consist in sampling the network
from a limited set of sources by using \texttt{traceroute} probes.
This methodology, akin to the merging of spanning trees from the 
different sources to a set of destinations,
leads necessarily to a partial, incomplete map of the Internet.
Accordingly, determination of Internet topology characteristics
from such sampled maps is in part a problem of statistical inference.
Our contribution 
begins with the observation that the
inference of many of the most basic topological quantities -- 
including network size and degree characteristics -- from
\texttt{traceroute} measurements is in fact a version of the
so-called `species problem' in statistics.  This observation has
important implications, since species problems are often quite 
challenging.  We focus here on the most fundamental
example of a \texttt{traceroute} internet species: the number
of nodes in a network.  Specifically, we characterize the difficulty
of estimating this quantity through a set of analytical arguments,
we use statistical subsampling principles to derive two 
proposed estimators, and we illustrate the performance of these
estimators on networks with various topological characteristics.
\end{abstract}


\section{Introduction}

A significant research and technical challenge in the study of large
information networks is related to the incomplete character of the
corresponding maps, usually obtained through some sampling process.  A
prototypical example of this situation is faced in the case of the
physical Internet. The topology of the Internet can be investigated at
different granularity levels such as the router and Autonomous System
(AS) level, with the final aim of obtaining an abstract representation
where the set of routers (ASs) and their physical connections (peering
relations) are the vertices and edges of a graph, respectively
\cite{mdbook,psvbook}. In the
absence of accurate maps, researchers rely on a general strategy that
consists in acquiring local views of the network from several vantage
points and merging these views.  Such local views are obtained by
evaluating a certain number of paths to different destinations, through
the use of probes or the analysis of routing tables, which we will
refer to generically in this paper as `\texttt{traceroute}-like sampling',
after the quintissential example of the well-known \texttt{traceroute}
tool. The merging of several of these views provides a map of a 
sampling of the Internet. 

While the knowledge of basic Internet topology (i.e., nodes and links)
discovered through such sampling is of significant value in and of itself,
it is natural to also want to use the
resulting sample maps to infer properties of the 
overall Internet map.  With such a strategy in mind, a number of research
groups have generated sample maps of the 
Internet~\cite{nlanr,caida,asdata,scan,lucent} that have 
then been used for the characterization of network properties.
For example, the `small world' character of the Internet has thus been
uncovered.  Moreover, the probability that any vertex in the 
graph has degree $k$ (i.e., that it has exactly $k$ links joining it to
immediate neighbors) has been characterized as being
skewed and heavy-tailed, with an approximately
power-law functional form~\cite{faloutsos}.

Recently, the question of the accuracy of the topological
characteristics inferred from such maps has been the subject of
various
studies~\cite{crovella,clauset,delos,Dallasta,achlioptas,latapy}.
Overall, these studies suggest that at a qualitative level the main
conclusions drawn from \texttt{traceroute}-like samplings are
reliable.  For example, it has been found that such samplings allow
for accurate discrimination between topologies with degree
distributions that are heavy-tailed from those that are
homogeneous~\cite{Dallasta}.  On the other hand, at a {\em quantitative}
level the evidence suggests the possibility for considerable
deviations between numerical summaries of characteristics of the
sampled networks and those of the actual Internet. 

The point of departure for our contributions 
is the observation 
that the inference, from \texttt{traceroute}-like
measurements, of many common measures of network graph characteristics is 
in fact related to the so-called `species problem' in statistics.
This association with the species problem has important implications
because, while the species problem is well-studied, it is also known to be
a statistical inference problem that is often particularly difficult.
Therefore, for example, in the context of Internet mapping and inference
with \texttt{traceroute}, while it is clear that the observed 
number of nodes, links, and vertex degrees necessarily will underestimate
the actual Internet values, it turns out that the accurate adjustment
of the observed values may be nontrivial.  Furthermore, the unique
nature of \texttt{traceroute}-like sampling procedures means that
standard tools for species estimation are unlikely to be immediately
applicable.

This paper is organized as follows. We provide general background
on \texttt{traceroute} and the species problem in Section~\ref{sec:bg}.
We then focus on what is arguably the most fundamental species
problem in the context of \texttt{traceroute}-like measurements:
inferring the number of nodes in a network.  In
Section~\ref{sec:analytic}, we present an analytical argument 
characterizing the structural elements relevant to this estimation 
problem.  In Section~\ref{sec:estimation}, we propose two estimators,
derived from principles of statistical subsampling.
In Section~\ref{sec:exp}, we describe the results of an extensive
numerical evaluation of these estimators.  Finally,
Section~\ref{sec:disc} contains some additional discussions and directions
for future work.

\section{Background}
\label{sec:bg}

\begin{figure}[thb]
\begin{center}
\includegraphics[width=10cm]{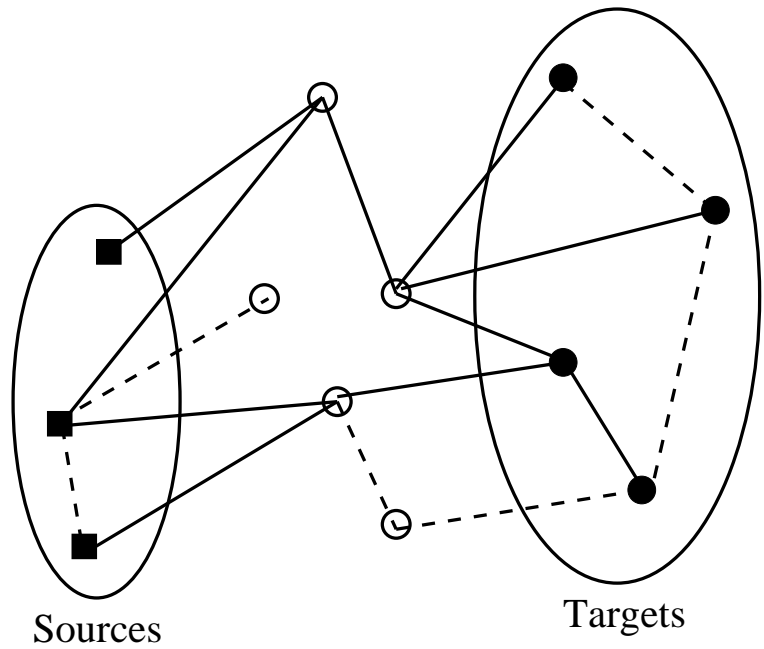}
\end{center}
\caption{Illustration of the \texttt{traceroute}-like procedure. 
Shortest paths between the set of sources and the set of destination 
targets are
discovered (shown in full lines) while other edges are not found
(dashed lines). In the case of degenerate shortest paths, only one
is found.} 
\label{fig:1}
\end{figure}

Throughout this paper we will represent an arbitrary network 
of interest as an undirected, connected graph $\G=(\V,\E)$,
where $\V$ is a set of vertices (nodes)
and $\E$ is a set of edges (links).  Denote by $N=|\V|$ and $M=|\E|$
the numbers of vertices and edges, respectively.
In a typical \texttt{traceroute} study, a set 
$S=\{s_1,\ldots,s_{n_S}\}$ of $n_S$ active sources
deployed in the network sends probes to a set 
$T=\{t_1,\ldots,t_{n_T}\}$ of $n_T$ destinations (or targets), 
for $S,T\subset \V$.
Each probe collects information on all the
vertices and edges traversed along the path connecting a source to
a destination~\cite{burch99}. 
The actual paths followed by the probes depend on many different
factors, such as commercial agreements, traffic congestion, and
administrative routing policies, but to a first approximation
are often thought of (and frequently modeled as) `shortest' paths.
The merging of the various sampled paths
yields a partial map of the network (Fig.\ref{fig:1}). 
This map may in turn be represented as a sampled subgraph 
$\G^*=(\V^*,\E^*)$.  

Numerous metrics are used in networking (and
indeed across the network-oriented sciences more generally) to summarize
characteristics of a network graph $\G$.  Some of the most fundamental
metrics include the number of vertices, $N$, the number of edges, $M$,
and the degrees $\{k_i\}$ of vertices $i\in \V$.  Many other metrics
either may be expressed as explicit functions of these or have
closely related behavior.  For an arbitrary metric, say 
$\eta\equiv \eta(\G)$, summarizing some characteristic of $\G$,
and a traceroute-sampled graph $\G^*$, 
it is natural to wish to produce an estimate, say $\hat\eta$
from the measurements underlying $\G^*$. 
However, some caution is in order, in that for the quantities $N$, $M$,
and $k_i$, the problem of their inference is closely related to
the so-called {\it species problem} in statistics.  

Stated generically,
the species problem refers to the situation in which, having observed
$n$ members of a (finite or infinite) population, each of whom falls
into one of $C$ distinct classes (or `species'), an estimate $\hat C$
of $C$ is desired.  This problem arises in numerous contexts, such
as numismatics (e.g., how many of an ancient coin were 
minted~\cite{esty}), linguistics (e.g., what was the size of
an author's apparent vocabulary~\cite{mcneil,et}), 
and biology (e.g., how many species of animals inhabit a given region).

The species problem has received a good deal of attention in statistics.
See~\cite{bungefitz} for an overview and an extensive 
bibliography.  Perhaps surprisingly, however, while the estimation
of the relative frequencies of species in a population is 
well-understood (given knowledge of $C$), the estimation of $C$
itself is often difficult.  In essence, what is needed is to 
estimate the number of species {\it not} observed.  This task is
problematic due to the fact that it is precisely the
species present in relatively low proportions in the population 
that are expected to be missed, and there could be an arbitrarily 
large number of such species in arbitrarily low proportions.
Despite (or perhaps because of) the difficulty of the problem,
numerous methods have been proposed for its solution, differing
mainly in the assumptions regarding the nature of the population,
the type of sampling involved, and the statistical machinery used.

An understanding of the implications of the species problem on 
network topology inference is of critical importance.  
For example, we note that in traceroute-like sampling the
problem of estimating the number of vertices $N$ in a network graph $\G$
may be mapped to a species problem by considering each separate vertex
$i$ as a `species' and declaring a `member' of the species $i$ to have
been observed each time that $i$ is encountered on one of the $n=n_S\, n_T$
\texttt{traceroute} paths.  A similar argument shows that estimation 
of the number of edges $M$ too may be mapped to a species problem.
Finally, as in \cite{zhang05}, 
the problem of inferring the degree $k_i$ of a vertex $i$
from traceroute measurements can also be mapped to the species problem,
by letting all edges incident to $i$ constitute a species and declaring
a member of that species to have been observed every time one of those
edges is encountered.  Because the values $N$, $M$, and 
$\{k_i\}_{i\in \V}$ are both important in their own right and
bear important relations to other metrics of interest, it is logical
to focus upon the question of their inference.  In this paper, we 
concentrate on the inference of the first of these quantities, $N$.

\section{Inferring $N$: Characterization of the Problem}
\label{sec:analytic}

Before proceeding to the construction of estimators for $N$,
as we will do in Section~\ref{sec:estimation}, it is useful 
to first better understand the structural elements of the problem.
In particular, the following analysis provides insight into 
the structure of the underlying `population', the relative frequency
of the various `species', and the impact of these factors on the problem
of inferring $N$.  For the sake of exposition, in this section we
adopt the common convention of modeling Internet routing, 
to a first approximation, as `shortest-path' routing.
However, we hasten to note that such an assumption, 
or even an assumption of a static routing protocol, are nowhere made
in the derivation of the estimators in Section~\ref{sec:estimation}.

A crucial quantity in the characterization of traceroute-like sampling
is the so-called betweenness centrality, which essentially counts for
each vertex the number of shortest paths on which it lies: nodes with
large betweenness lie on many shortest paths and are thus more easily
and more frequently probed~\cite{Dallasta}. More precisely, if ${\cal
D}_{hj}$ is the total number of shortest paths from vertex $h$ to
vertex $j$, and ${\cal D}_{hj}(i)$ is the number of these shortest
paths that pass through the vertex $i$, the betweenness of the vertex
$i$ is defined as $b_i=\sum {\cal D}_{hj}(i)/{\cal D}_{hj}$, where the
sum runs over all $h,j$ pairs with $j \neq h\neq i$.  
It can be shown~\cite{Goh:2003} that the average shortest path
length between pairs of vertices, $\ell$, is related to the betweenness
centralities through the expression
$$
\sum_i b_i = N(N-1)(\ell -1) \ .
$$
This may be rewritten in the form
\begin{equation}
N= 1 + \frac{E[b]}{\ell -1} \ ,
\label{eq:Nandaveb}
\end{equation}
where the expectation $E[\cdot]$ is with respect to the
distribution of betweenness across nodes in the network i.e.,
$P(b)=\#\{i\in \V\, :\, b_i=b\}/N$.

Empirical experiments suggest that the average
shortest path length $\ell$ can be estimated quite accurately,
which is not surprising given the path-based nature of 
\texttt{traceroute}.  Therefore, the problem of estimating $N$
is essentially equivalent to that of estimating the
average betweenness centrality.  Motivated by the fact that
Internet maps have been found to display a broad distribution of
not only degrees, but also betweenness~\cite{Dallasta},
let us consider a model that pictures the
distribution of the betweenness  as divided into
two parts.  That is, we model the distribution $P(b)$ as
a mixture distribution~\cite{mclachlan_peel}
\begin{equation}
P(b) = \pi P_1(b) + (1-\pi) P_2(b) ,
\end{equation}
where $P_1$ is a distribution at low values $b\in [1, b_{min})$,
for some $b_{min}$ small, and $P_2(b)$ is a distribution
at high values
$b\in [b_{min},b_{max}]$, $b_{max}>>b_{min}$.

The average $E[{b}]$ in (\ref{eq:Nandaveb}) is a weighted combination 
of two terms i.e., $E[{b}]  = \pi E_1[b] +(1-\pi) E_2[b] $.  
From the perspective of the simple parametric model just described, 
the challenge of accurately estimating $E[b]$ --
and hence $N$ -- can be viewed as 
a problem of the accurate estimation
of the two means, $E_1[b]$ and $E_2[b]$, and the weight $\pi$.
Unfortunately, the first mean, $E_1[b]$, requires knowledge of the
betweenness of vertices with ``small'' betweenness.  That is,
knowledge of nodes $i\in \V$ traversed by relatively few paths.
But these are precisely the nodes on which we receive the least
information from \texttt{traceroute}-like studies, as they are
expected to be visited infrequently or not at all.  And the
relative proportion $\pi$ of such nodes would seem to be similarly
difficult to determine.  As mentioned earlier, this is a hallmark
characteristic of the species problem, i.e. the lack of accurate
knowledge of the relative number in the population of comparitively
infrequently observed species.

As for the second mean, $E_2$, let us approximate the observed
broad distribution of betweeness in the tail by a heavy-tailed
power-law form i.e., $P_2(b)=b^{-\beta}/K$, where $K$ is a 
normalization constant.  Then 
\begin{equation}
\label{eq:E2}
E_2[{b}] =
\frac{1}{K} \int_{{b}_{min}}^{{b}_{max}}
{b}^{1-\beta} d{b} \ .
\end{equation}
A simple calculation yields
$K = ({b}_{max}^{1-\beta}-{b}_{min}^{1-\beta})/(1-\beta)$.  Additionally,
if the only origin of the cutoff is the finite size of the network,
${b}_{max}$ can be defined by imposing the condition that the expected 
number of nodes beyond the cut-off is bounded by a fixed constant~\cite{mdbook}.
Therefore one finds that
\begin{equation}
\label{eq:bmax}
N \times \int_{{b}_{max}}^\infty P({b}) d{b} \sim 1
\quad\Rightarrow\quad
{b}_{max} \sim \left(
\frac{(\beta -1) K }{(1-\pi) N} \right)^{\frac{1}{1-\beta}}
\sim {b}_{min}((1-\pi) N)^{-\frac{1}{1-\beta}}
 ,
\end{equation}
i.e. a relation between  ${b}_{max}$,  ${b}_{min}$
$N$, $\pi$ and $\beta$, in which we have also
used the assumption ${b}_{min} \ll {b}_{max}$ that
implies $K \sim {b}_{min}^{1-\beta}/(\beta-1)$.

Our empirical studies indicate that the exponent $\beta$ can
be estimated fairly accurately from the distribution of betweenness'
observed through \texttt{traceroute} measurements.  And the above
calculations suggest that knowledge of $\beta$ is key to knowledge of $K$.
However, note from (\ref{eq:bmax}) that $b_{max}$ involves not only
the unknown $N$, as would be expected, but also $\pi$, which suggests
that even the inference of the $E_2$ component of $E[b]$ is
potentially impacted by our ability (or lack thereof) to recover
information on nodes with low betweenness.  Furthermore, we mention
that our numerical studies show that $\beta$ is in fact likely 
quite close to $2$ in the real Internet, which suggests an additional
level of subtlety in the accurate estimation of $E_2$, due to the
nature of the integral in (\ref{eq:E2}).

The above analysis both highlights the relevant aspects of the
species problem inherent in estimating $N$ and indicates the
futility of attempting a classical parametric estimation approach.
One is led, therefore, to consider nonparametric methods, in which
models with a small, fixed number of parameters are eschewed in 
favor of models that essentially have as many parameters as data.

From the perspective of classical nonparametric species models
in statistics, the 
estimation of the total number of vertices $N$, the total number of edges 
$M$, and the node degree $k_i$ are all non-standard statistical inference 
problems. Consider the classical idealized model where the observed frequencies 
for different species are truncated Poisson variables conditionally on their 
positivity. Suppose the Poisson intensities for all the species (including 
unobserved ones) form a random sample from a completely unknown distribution. 
Then it is known that in this nonparametric Poisson mixture model, 
the estimation 
of the total intensity of unobserved species is a well-posed problem 
\cite{good,robbinszhang}, but the estimation of the total number of species is 
ill-posed \cite{zhang05} from an information theoretical point of view. 
This indicates the ill-posedness of the problems of estimating $M$ and $k_i$ 
without assuming a parametric model for the distribution of the betweenness 
centrality, since under Poissonized sampling, the betweenness centrality is 
proportional to the marginal intensity for links, or species in these problems. 
However, for the estimation of $N$, vertices are treated as species, and they 
can be thought of as being first sampled with roughly equal probability 
as targets and then with unequal 
probability as intermediate nodes in \texttt{traceroute} experiments. 
This suggests the estimation of $N$ is more akin to that of the 
total intensity of unobserved species, since the total unobserved intensity 
is simply the product of the
number of unobserved species and the common intensity when the species are 
equally likely to be included in the sample. This observation is
crucial in our derivation of the leave-one-out estimator in 
Section~\ref{sec:leave-one-out}. 

\section{Estimators of Network Size}
\label{sec:estimation}

A naive estimator of $N$ is simply $N^*$, the number of nodes observed
in the \texttt{traceroute} study.  Given the levels of 
coverage afforded by the scale of current Internet mapping initiatives,
$N^*$ can be expected to vastly underestimate $N$ (e.g., \cite{Dallasta}).
Motivated by the results and discussion in Section~\ref{sec:analytic},
in this section we develop two nonparametric estimators for $N$, using
subsampling principles.

\subsection{A Resampling Estimator}
\label{sec:resampling}

A popular method of subsampling is that of resampling, which
underlies the well-known `bootstrap' method \cite{efron}.  
Given a sample $x_1,...,x_n$ from a population, resampling in
its simplest form means taking a second sample $x^*_1,...,x^*_m$
from $x_1,...,x_n$ to study a certain relationship between the first
sample and the population through the observed relationship
between the second and first samples.
We utilize a similar principle here to obtain a
factor by which the observed number of vertices $N^*$ is inflated
to yield an estimator $\hat N_{RS}$ of $N$. 

Consider the quantity $N^*/N$ i.e., the fraction of nodes discovered
through \texttt{traceroute} sampling of $\G$, which we will call the
{\it discovery ratio.}  The expected discovery ratio $E[N^*/N]$ has been
found to vary smoothly as
a function of the fraction $q_T=n_T/N$ of targets sampled, for a given
number $n_S$ of sources~\cite{Dallasta,latapy}.  We will use this
fact, paired with an assumption of a type of scaling relation on $\G$,
to construct our estimator for $N$.  Specifically, we will assume that
the sampled subgraph $\G^*$ is sufficiently representative of $\G$
so that a sampling ratio on $\G^*$ similar to that used in its obtention
from $\G$ yields a discovery ratio similar to the fraction of nodes
discovered in $\G$.  

\begin{figure}[thb]
\centering
\includegraphics[width=14cm]{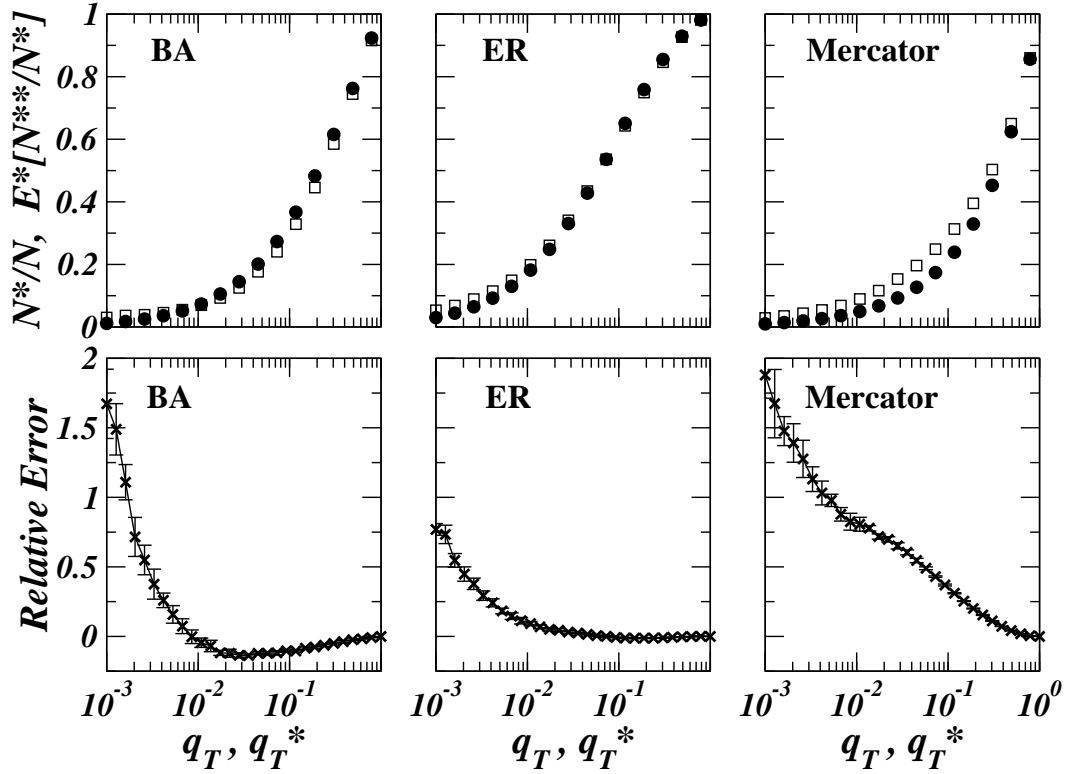}
\caption{A comparison of the quantities $N^*/N$ and 
$E^*[N^{**}]/N^*=E[N^{**}|G^*]/N^*$,
as a function respectively of $q_T=n_T/N$ and $q^*_T=n^*_T/N^*$,
for the three networks described in Section~\ref{sec:exp}.
Here $q^*_T=q_T$ and $n_S=n_S^*=10$.
Top row shows the averages of $N^*/N$ and $E^*[N^{**}]/N^*$ 
over 100 realizations
of $G^*$.  Bottom row shows the average of the difference of these two
quantities, relative to $N^*/N$, over the same 10 realizations.
The comparison in the top row confirms the validity of the scaling assumption
underlying the resampling estimator derived in Section~\ref{sec:resampling},
while the comparison in the bottom row indicates better performance
of the estimator can be expected with increasing $q_T$.  (Note: One
standard deviation error bars are smaller than the symbol size in most cases.)}
\label{valid_resampling}
\end{figure}

That is, suppose that we choose a set $S^*$ of $n^*_S$ source vertices
in $\G^*$ and a set $T^*$ of $n^*_T$ target vertices, in a manner
similar to the way that the original sets $S$ and $T$ underlying $\G^*$
were chosen, and such that $q^*_S\sim q_S$ and $q^*_T\sim q_T$,
where $q^*_S=n^*_S/N^*$, $q^*_T = n^*_T/N^*$, $q_S=n_S/N$ and $q_T$  
is defined above.  Then we assume that the result of a
\texttt{traceroute} study on $\G^*$, from sources in $S^*$ to targets
in $T^*$, will yield a subsubgraph, say $\G^{**}$, of $N^{**}$ nodes,
such that on average the discovery ratio $N^{**}/N^*$ on $\G^*$ is similar to 
the fraction of vertices of $\G$ discovered originally through $\G^*$.
In other words, we assume that $N^*/N \sim E[N^{**}/N^*\,|\,\G^*]$, where
the expectation $E[\,\cdot\,|\G^*]$ is with respect to whatever 
random mechanism drives the choice of source and target sets 
$S^*$ and $T^*$ on $\G^*$, conditional on fixed $\G^*$. 
Our empirical studies, using uniform random sampling on the
networks described in Section~\ref{sec:exp}, suggest that
this assumption is quite reasonable over a broad range of values
for $q_T$, as shown in Fig.~\ref{valid_resampling}.

Writing $E^*[\cdot]\equiv E[\,\cdot\,|\G^*]$,
the condition of equal discovery rates can be rewritten 
in the form $N\sim N^* (N^*/E^*[N^{**}])$.  The quantity 
$E^*[N^{**}]$ can be estimated by repeating the resampling experiment
just described some number $B$ of times, compiling subsubgraphs
$\G^{**}_1,\ldots,\G^{**}_B$ of sizes $N^{**}_1,\ldots,N^{**}_B$,
and forming the 
average $\bar{N}^{**} = (1/B)\sum_k N^{**}_k$.
Substitution then yields 
\begin{equation}
\label{eq:NhatRS}
\hat{N}_{RS} = N^*\,\cdot\, \frac{N^*}{\bar{N}^{**}}
\end{equation}
as a resampling-based estimator for $N$.  

\begin{figure}[thb]
\centering
\includegraphics[width=12cm]{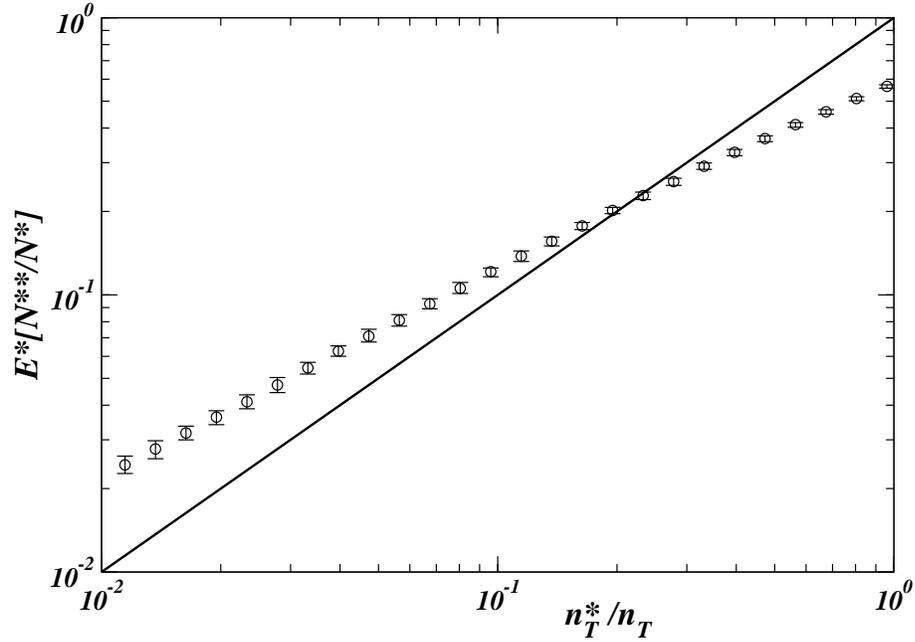}
\caption{Illustration of the obtention of the resampling
estimator, in the case of a BA graph of size $N=10^5$. The initial
sampling was obtained with $n_S=10$ sources and $n_T=10^4$ targets
($q_T=0.1$), yielding a graph $\G^*$ of size $N^*=33178$
(and $M^*=133344$). The circles show the ratio of the average
size of the resampled graph $\G^{**}$, $\bar{N}^{**}/N^*$, 
as a function of the ratio $n_T^*/n_T$, with $n_S^*=n_S=10$ sources.
The errorbars give the variance with respect to the various
placements of sources and targets used for the resampling.
The straight line is $y=x$ and allows to find the 
value of $n_T^*$ such that $n_T/n_T^*=N^*/\bar{N}^{**}$
}
\label{fig:inters}
\end{figure}

Note, however, that 
its derivation is based upon the premise that $q^*_S=q_S$ and $q^*_T=q_T$, 
and $q_S, q_T$ are unknown (i.e., since $N$ is unknown).  To address
this issue, we first let $n^*_S=n_S$, since typically the number of sources
is too small to make $q_S$ a useful quantity.  Then we 
note that the expression $q^*_T=q_T$, in conjunction
with our assumption on discovery rates, together imply that
$n^*_T/n_T \sim E^*[N^{**}]/N^*$.  With respect
to the calculation of $\hat{N}_{RS}$, this fact suggests the strategy
of iteratively adjusting $n^*_T$ until the relation
$n^*_T/n_T\approx \bar{N}^{**}/N^*$ holds. 
Alternatively, one may picture the situation geometrically, as 
shown in Fig.~\ref{fig:inters}.
The value of $\bar{N}^{**}$
for the appropriate $n^*_T$ is then substituted into 
(\ref{eq:NhatRS}) to produce $\hat{N}_{RS}$. 
In practice, one may either use a fixed value of $B$ throughout or,
as we have done, increase $B$ as the algorithm approaches
the condition $n^*_T/n_T\approx \bar{N}^{**}/N^*$.

\subsection{A `Leave-One-Out' Estimator}
\label{sec:leave-one-out}

Various other subsampling paradigms might be used to
construct an estimator. A popular one is the `leave-one-out'
strategy underlying such methods as `jack-knifing' 
\cite{quenouille,tukey} 
and `cross-validation'~\cite{efron}, which amounts to subsampling $\G^*$ with 
$n^*_T=n_T-1$.  The same underlying principle may be applied in a useful
manner to the problem of estimating $N$, in a way that 
does not require the scaling assumption underlying (\ref{eq:NhatRS}),
as we now describe.

Recall that $\V^*$ is the set of all vertices discovered by a
\texttt{traceroute} study, including the $n_S$ sources 
$S=\{s_1,\ldots,s_{n_S}\}$ and the $n_T$ targets 
$T=\{t_1,\ldots,t_{n_T}\}$.  Our approach will be to connect
$N$ to the frequency with which individual targets $t_j$ are included
in traces from the sources in $S$ to the other targets in 
$T\setminus \{t_j\}$.  Accordingly, let $\V^*_{i,j}$ be the
set of vertices discovered on the path from source $s_i$ to target
$t_j$, inclusive of $s_i$ and $t_j$.  Then the set of vertices
discovered as a result of targets other than a given $t_j$ can be
represented as $\V^*_{(-j)} = \cup_i\cup_{j'\ne j} \V^*_{i,j'}$.
Next define $\delta_j = I\left\{t_j\notin \V^*_{(-j)}\right\}$ 
to be the indicator
of the event that target $t_j$ is {\em not} `discovered' by traces to
any other target.  The total number of such targets 
is $X=\sum_j \delta_j$.

We will derive a relation between $X$ and $N$ through consideration of
the expectation of the former. Under an assumption of simple
random sampling in selecting target nodes from $\V$, 
given a pre-selected (either randomly or not) set of source nodes, we have
\begin{equation}
\Pr\left(\delta_j=1\,|\, \V^*_{(-j)}\right) = 
\frac{N - N^*_{(-j)}}{N-n_s-n_T+1} \enskip ,
\label{eq:dgivenV}
\end{equation}
where $N^*_{(-j)}=\left|\V^*_{(-j)}\right|$.  Note that, by symmetry,
the expectation $E\left[N^*_{(-j)}\right]$ is the same for all $j$: we denote
this quantity by $E\left[N^*_{(-)}\right]$.  As a result of these two facts, 
we may write
\begin{equation}
\label{eq:E[X]}
E[X] = \sum_j \frac{N - E\left[N^*_{(-j)}\right]}{N-n_s-n_T+1} 
     = \frac{n_T\left(N - E\left[N^*_{(-)}\right]\right)}{N-n_s-n_T+1}
\enskip ,
\end{equation}
which may be rewritten as
\begin{equation}
\label{eq:Nl1out}
N = \frac{n_T E\left[N^*_{(-)}\right] - (n_s+n_T-1) E[X]}{n_T - E[X]}
\enskip .
\end{equation}

To obtain an estimator for $N$ from this expression it is necessary to
estimate $E\left[N^*_{(-)}\right]$ and $E[X]$, for which it is natural
to use the unbiased estimators $\bar{N}^*_{(-)}= (1/n_T)\sum_j
N^*_{(-j)}$ and $X$ itself, which is measured during the 
\texttt{traceroute} study.  However, while substitution of these
quantities in the numerator of (\ref{eq:Nl1out}) is fine, substitution
of $X$ for $E[X]$ in the denominator can be problematic in the event
that $X=n_T$. Indeed, when none of the targets $t_j$ are discovered
by traces to other targets, as is possible if $q_T=n_T/N$ is
small, $N$ will be estimated by infinity. A better strategy is to
estimate the quantity $1/(n_T-X)$ directly. Under the condition that
$N^*_{(-j)}\approx N^*_{(-j')}\approx N^*_{(-j,-j')}$, where
$N^*_{(-j,-j')}=\left|\V^*_{(-j)}\cap \V^*_{(-j')}\right|$, and
our assumption of simple random sampling of 
target vertices,
it is possible to
produce an approximately unbiased estimator of this quantity, which
upon substitution yields
\begin{equation}
\label{eq:NhatL1O}
\hat{N}_{L1O} = \frac{n_T+1}{n_T}\,\cdot\,
                \frac{n_T \bar{N}^*_{(-)} - (n_s+n_T-1) X}{n_T+1 - E[X]}
\enskip .
\end{equation}

Formal derivation of the leave-one-out estimator $\hat{N}_{L1O}$
in (\ref{eq:NhatL1O}) may be found in the appendix.  Note that
even if $X=n_T$, the estimator remains well-defined.  The condition
that all $\V^*_{(-j)}$ and their pairwise intersections have approximately 
the same cardinality is equivalent to saying that the unique contribution
of discovered vertices by any one or any pair of vertices is relatively
small.  For example, using data collected by the Skitter project at 
CAIDA \cite{caida}, a fairly uniform discovery rate of roughly 
$3$ new nodes per new target, after the initial $200$ targets, 
has been cited~\cite{marginal}.  We have found too that a similar rate
held in the empirical experiments of Section~\ref{sec:exp}.
Note that this condition also implies that $N^*_{(-j)}\approx N^*$, for 
all $j$, which suggests replacement of $\bar{N}^*_{(-)}$ by $N^*$
in (\ref{eq:NhatL1O}).  Upon doing so, and after a bit of algebra,
we arrive at the approximation
\begin{equation}
\label{eq:NhatL1Ofinal}
\hat{N}_{L1O}\approx (n_S+n_T) \,+\, \frac{N^* - (n_S+n_T)}{1-w^*}
\enskip ,
\end{equation}
where $w^* = X/(n_T+1)$, $X$ being the number of targets not
discovered by traces to any other target.

In other words, $\hat{N}_{L1O}$ can be seen as
counting the $n_S+n_T$ vertices in $S\cup T$ separately, and then
taking the remaining $N^*-(n_S+n_T)$ nodes that were `discovered'
by traces and adjusting that number upward by a factor of $(1-w^*)^{-1}$.
This form is in fact analogous to that of a classical method in the
literature on species problems, due to 
Good \cite{good}, in which the
observed number of species is adjusted upwards by a similar factor
that attempts to estimate the proportion of the overall population
for which no members of species were observed.  Such estimators
are typically referred to as coverage-based estimators, and a combination
of theoretical and numerical evidence seems to suggest that they enjoy
somewhat more success than most alternatives \cite{bungefitz}.

\section{Numerical Validation}
\label{sec:exp}

We examined the performance of the estimators proposed in 
Section~\ref{sec:estimation} using a methodology similar to
those in~\cite{Dallasta,clauset,latapy}.  That is, we began with
known graphs $\G$ with various topological characteristics,
equipped each with an assumed routing structure, performed
a \texttt{traceroute}-like sampling on them, which yielded
a sample graph $\G^*$, and computed the estimators
$\hat{N}_{RS}$ and $\hat{N}_{L1O}$.  This process was repeated
a number of times, for various choices of source and target nodes,
at each of a range of settings of the parameters 
$N$, $n_S$, and $n_T$.  A performance 
comparison was then made by comparing values of $\hat{N}/N$,
for $\hat{N}=N^*, \hat{N}_{RS}$, and $\hat{N}_{L1O}$.

\subsection{Design of the Numerical Experiments}

Three network topologies were used in our experiments, two synthetic and one
based on measurements of the real Internet.  The synthetic topologies
were generated according to (i) the classical Erd\"os-R\'enyi (ER) 
model~\cite{er} and (ii) the network growth model of 
Albert and Barab\'asi (BA)~\cite{sf}.  This choice of topologies allows
us to examine the effects of one of the most basic distinguishing
characteristics among networks, the nature of the underlying degree
distribution.  In particular, the ER model is the standard example 
of a class of {\em homogeneous} graphs, in which the the degree distribution
$P(k)$ has small fluctuations and a well defined average degree, while
the BA model is the original example of a class of 
{\em heterogeneous} graphs, for which $P(k)$ is a broad
distribution with heavy-tail and large fluctuations, spanning various
orders of magnitude. 
%
In our experiments, we have used randomly generated ER and BA 
networks with average degree $6$, and sizes $N$ ranging from $10^3$ 
to $10^6$ nodes.

The ER and BA models are standard choices for experiments like ours, and 
useful in allowing one to assess the effect on a proposed methodology
of a broad degree distribution, but they
lack other important characteristics of the real Internet, such
as clustering, complex hierarchies, etc.  Therefore, we used as our third
topology the Internet sample from MERCATOR
\cite{mercator}, a graph with 
$N=228,263$ nodes and $M=320,149$ edges.
While there are newer Internet graphs, such as those from 
CAIDA~\cite{caida}, our choice of MERCATOR is influenced by the
fact that it resulted from an attempt to have
obtained an exhaustive map of the Internet in 1999.  The aim in presenting
such results is primarily illustrative.


Given a graph $\G$, and a chosen set of values for $N$, $n_S$, and
$n_T$, a \texttt{traceroute}-like study was simulated as follows.
First, a set of $n_S$ sources $S=\{s_1,\ldots,s_{n_S}\}$ were
sampled uniformly at random from $\V$ and a set of $n_T$ targets 
$T=\{t_1,\ldots,t_{n_T}\}$ were sampled uniformly at random from 
$\V\setminus S$.  Second, paths from each source to all targets
were extracted from $\G$, and the merge of these paths was returned
as $\G^*$.  Shortest path routing, with respect to common edge 
weights $w_e\equiv 1$, was used in collecting these simulated 
\texttt{traceroute}-like data, based on standard algorithms.  
Unique shortest paths
were forced by breaking ties randomly.  Other choices of routing
between sources and targets, such as random shortest path and all
shortest paths, have been found to lead to similar behavior with
respect to discovery rates of nodes and links~\cite{Dallasta}.  After
initial determination, routes are considered fixed, in that the route 
between a source $u_s \in S$ and a vertex $v \in V$ is always
the same, independent of the destination target $u_t \in T$.


We note that the routing model used here is chosen simply as
a first approximation to that in the real Internet, and emphasize
that the estimators proposed in Section~\ref{sec:estimation} are not
derived in a manner that makes any explicit use of these routing 
assumptions.  This model has been used in a number of recent
papers~\cite{clauset,achlioptas,Dallasta,latapy} and, although it does not
account for all realistic subtleties, we have found, as in previous
studies, that it appears to be sufficient for studying the essence of the 
issues at hand regarding inferences of Internet topology `species'.
Further studies could incorporate refinements of the model, such as the
ones proposed in \cite{timur}.

\subsection{Results}

\begin{figure}[thb]
\centering
\includegraphics[width=15cm]{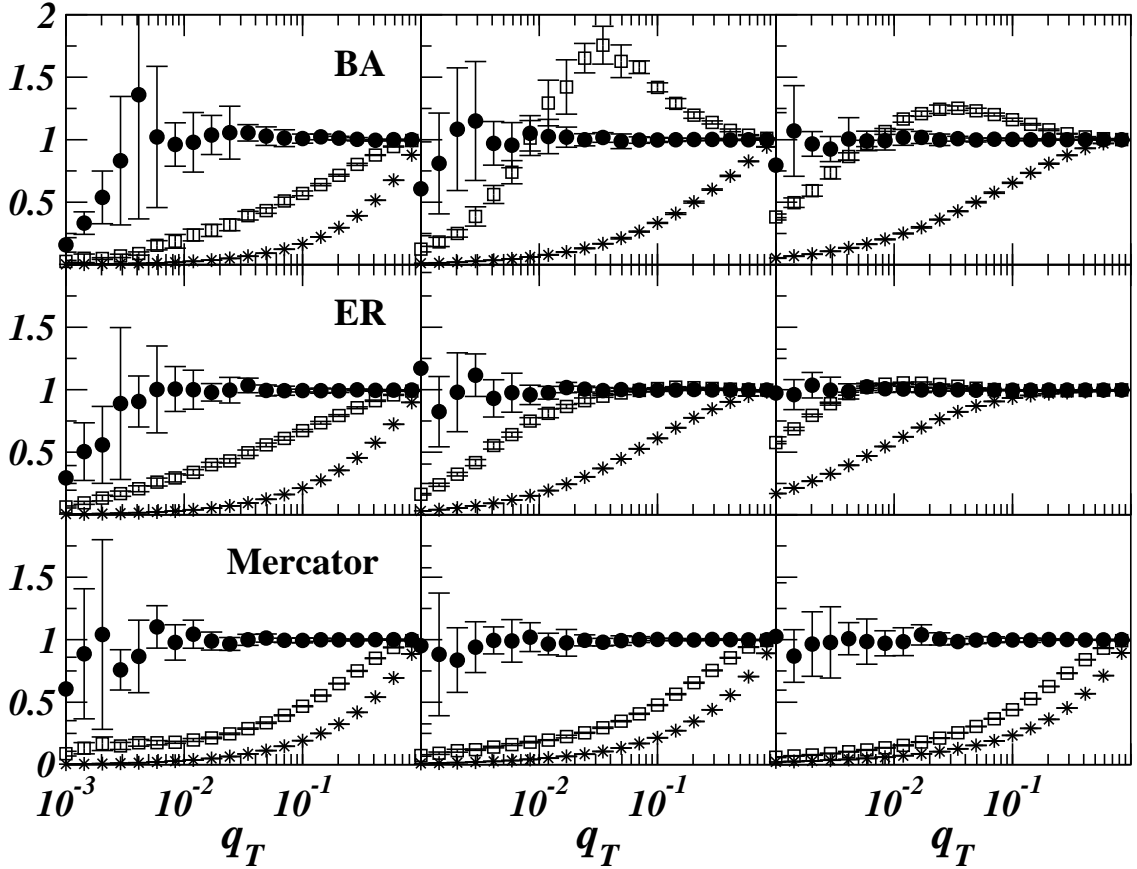}
\caption{Comparison of the various estimators for
the BA (top), ER (middle) and Mercator (bottom) networks.
The curves show the ratios of the various estimators
to the true network size, as a function of the target density $q_T$.
Full circles: $\hat{N}_{L1O}/N$ ; Empty squares:
$\hat{N}_{RS}/N$; Stars: $N^*/N$.  Values and one standard deviation
error bars are based on 100 trials, with random choice of 
sources and targets for each trial.
Left figures: $n_S=1$ source; Middle: $n_S=10$ sources; Right:
$n_S=100$ sources.
}
\label{estimators}
\end{figure}

The plots in Fig.~\ref{estimators} show a comparison of
$N^*/N$, $\hat{N}_{RS}/N$, and $\hat{N}_{L1O}/N$, for
$n_S=1,10,$ and $100$ sources, as a function of $q_T$.
A value of $1$ for these ratios is desired, and it is clear that
in the case of both the resampling and the ``leave-one-out'' estimator
that the improvement over the ``trivial'' estimator $N^*$ is substantial. 
Increasing either the number of sources $n_S$ or the density of targets 
$q_T$ yields better results, even for $N^*$, but the estimators we propose
converge much faster than $N^*$ towards values close to the true size
$N$.  

Between the resampling and the ``leave-one-out'' estimator, the latter
appears to perform much better.  For example, we note that while both 
estimators suffer from a downward bias for very low values of $q_T$, this
bias persists into the moderate and, in some cases, even high range
for the resampling estimator.  
This is probably due to the fact that the basic hypothesis of scaling
underlying the derivation of $\hat{N}_{RS}$ is only approximately
satisfied, while for $\hat{N}_{L1O}$, the underlying hypotheses are
indeed well satisfied.
Notice, however, that the ``leave-one-out'' 
estimator has a larger variability at small values
of $q_T$, while that of the resampling estimator is fairly 
constant throughout.  This is because the same number $B$ of
resamples is used in calculating $\hat{N}_{RS}$ in equation 
(\ref{eq:NhatRS}), and the uncertainty can be expected to scale
similarly, but in calculating $\hat{N}_{L1O}$ in equation
(\ref{eq:NhatL1O}), the uncertainty will scale with 
$n_T$ (and hence $q_T$).  

In terms of topology, estimation of $N$ appears to be easiest
for the ER model.  Even $N^*$ is more accurate i.e., the discovery
rate is higher.  Estimation on the MERCATOR graph appears to be
the hardest, although interestingly, the performance of the 
``leave-one-out'' estimator 
seems to be approximately a function of $N^*/N$ and $n_T$ and thus 
quite stable in all three graphs. 
The MERCATOR graph has a much higher 
proportion of low-degree vertices than the two synthetic
graphs, which therefore have particularly small betweenness (and 
thus lie on very few shortest paths) and are very difficult to discover.
On a side note, we mention too that the
resampling estimator behaves in a rather curious, non-monotonic
fashion in two of the plots, as $q_T$ grows.  
At the moment, we do not have a reasonable explanation
for this behavior, although we note that it appears to be constrained
to the case of the BA graph and that some indication of this behavior
can already be seen for this graph in Fig.~\ref{valid_resampling}.

\begin{figure}[thb]
\centering
\includegraphics[width=14cm]{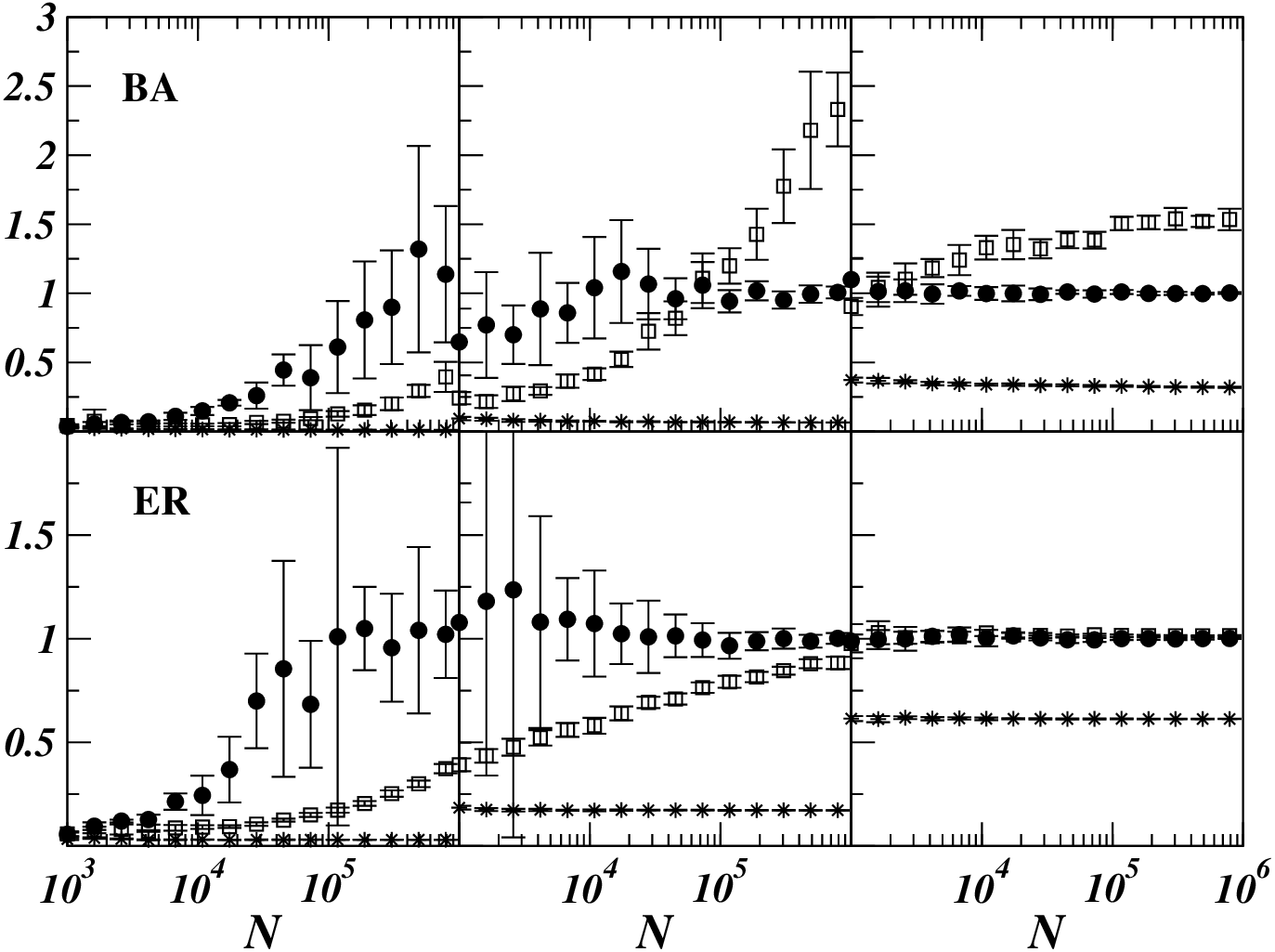}
\caption{Effect of the size $N$ of the graph $\G$ for BA and ER graphs
at constant number of sources and density of targets.
The curves show the ratios of the various estimators
to the true network size, as a function of the graph size $N$.
Full circles: $\hat{N}_{L1O}/N$ ; Empty squares:
$\hat{N}_{RS}/N$; Stars: $N^*/N$. 
Values and one standard deviation
error bars are based on 100 trials, with random choice of
sources and targets for each trial.
$n_S=10$.
Left figures: $q_T=10^{-3}$; Middle: $q_T=10^{-2}$; Right:
$q_T=10^{-1}$.
}
\label{effectsize}
\end{figure}

\begin{figure}[thb]
\centering
\includegraphics[width=14cm]{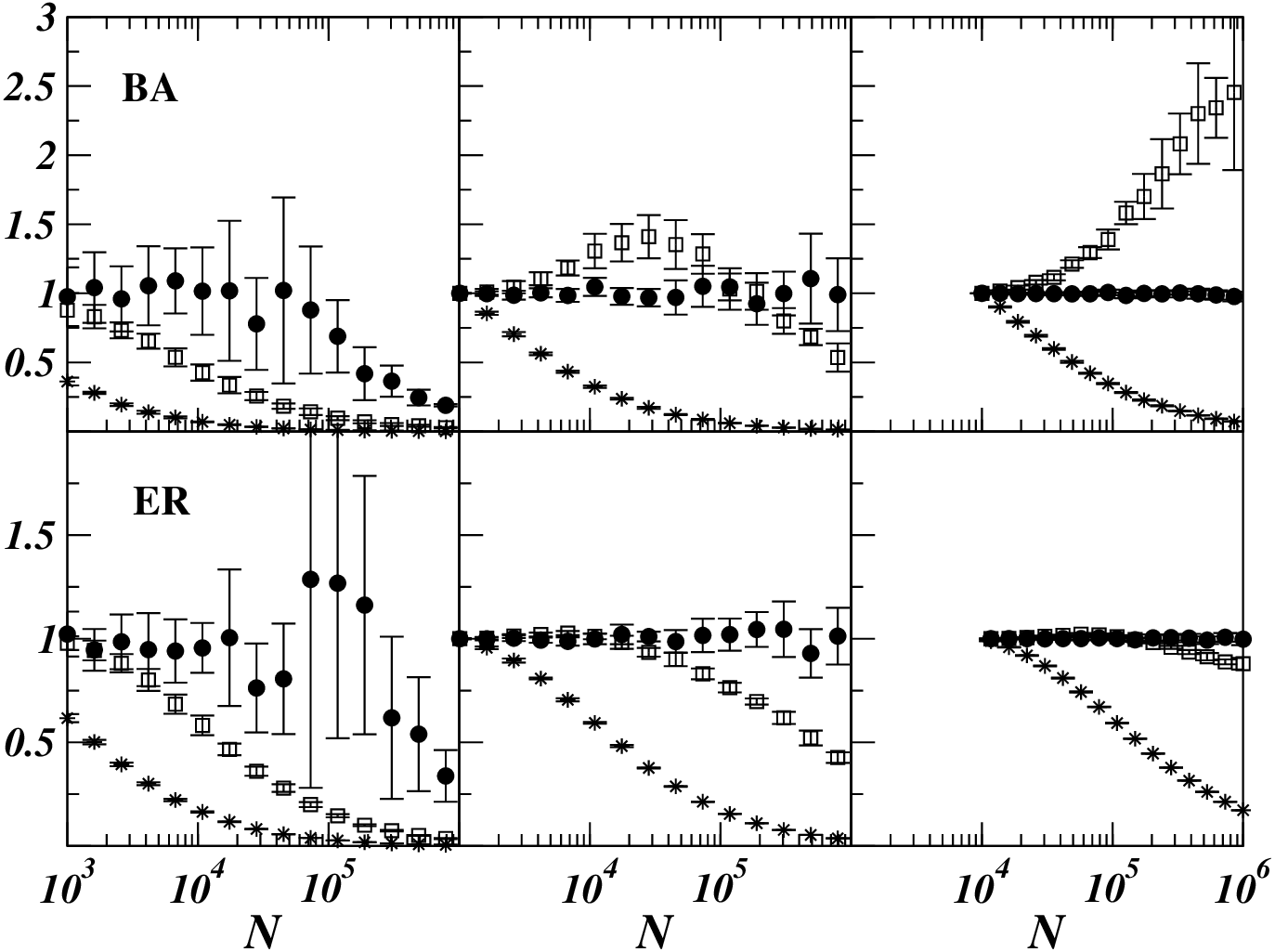}
\caption{Effect of the size $N$ of the graph $\G$ for BA and ER graphs
at constant number of sources and targets.
The curves show the ratios of the various estimators
to the true network size, as a function of the graph size $N$.
Full circles: $\hat{N}_{L1O}/N$ ; Empty squares:
$\hat{N}_{RS}/N$; Stars: $N^*/N$. 
Values and one standard deviation
error bars are based on 100 trials, with random choice of
sources and targets for each trial.  $n_S=10$.
Left figures: $n_T=10^{2}$ targets; Middle: $n_T=10^3$ targets; Right:
$n_T=10^{4}$ targets.
}
\label{effectsizebis}
\end{figure}

In Fig. \ref{effectsize}, we investigate, at fixed $n_S$ and $q_T$, the 
effect of the real size of the graph $N$. Interestingly, the estimators 
perform better for larger sizes, while $N^*/N$ on the contrary decreases.
This is due to the fact that the sample graph $\G^*$ gets bigger, providing
more and richer information, even if the discovery ratio does not grow.
The odd nature of the results for the BA graph comes from the peak
associated with the resampling estimator mentioned earlier; 
see Fig.~\ref{estimators}.
At a fixed number $n_T$ of targets, however, the quality of
the estimators $\hat{N}_{RS}$ and $\hat{N}_{L1O}$ gets worse 
as $N$ increases, as shown in Fig. \ref{effectsizebis}.

\section{Discussion}
\label{sec:disc}

In this paper, we have investigated the problem of inferring 
a network's properties from \texttt{traceroute}-like measurements
in the framework of the so-called `species' problem. As a first example
of application, we have focused on the issue of estimating the
real size $N$ of a network from only the knowledge of the sampled
graph.  Despite the fact that species problems often can be quite
difficult, in this case we find it is possible to propose
an estimator that, based on our empirical studies, works quite well,
even at quite low sampling densities.

While the present study provides a first promising step that clearly
illustrates the relevance of the species problem in Internet inference,
numerous issues remain to be explored, even simply in the case of
estimation of $N$ alone.  For example, the proposed
estimators could be evaluated with other types of networks. 
Similarly, one could examine the effect of  non-random source 
placement (e.g., restricted only to the fringe of the network), 
as well as that of more realistic \texttt{traceroute} models
\cite{timur}.  However, in the case of these latter two changes,
we would not expect the `leave-one-out' estimator to suffer
much in performance, since its derivation assumes only uniform 
random choice of targets, and not sources, and furthermore 
makes no explicit assumptions about routing.
The effect of the inclusion in $\G^*$ of the paths from each
source to the other sources should as well be investigated.

Our results showed that the `leave-one-out' estimator performed
noticeably better than the resampling estimator.  Nevertheless,
the resampling estimator should not be summarily dismissed quite yet.
In particular, while the derivation of the `leave-one-out' estimator
is quite specific to the problem of estimating $N$, the derivation of the
resampling estimator is general and independent of what is to be
estimated.  Initial experiments indicate that in estimating
the number of edges $M$ in a network, for example, a resampling
estimator yields similar improvements over the observed
value $M^*=|E^*|$ as seen in estimating $N$.  On the other hand, it
is not immediately apparent how the leave-one-out principle might
be applied to estimating $M$, as it is nodes (i.e, targets) and not edges 
that are chosen at the start of a \texttt{traceroute} study.

Finally, it is worth recalling the broader issue raised by this paper:
the fact that the problem of estimating a characteristic $\eta(\G)$
of a network graph $\G$, based on a sampled subgraph $\G^*$, is as
yet poorly understood.  We have taken the case of $\eta(\G)=N$ as
a prototype to explore and illustrate.  However, for this case
alone there are natural alternatives that one might consider.  For
example, an experiment that used \texttt{ping} to test for the response 
of some sufficient number $n$ of randomly chosen IP addresses could
yield an estimator $\hat{a}$ of the fraction of `alive' addresses and,
in turn, an estimator $\hat{N}_{ping}=2^{32}\hat{a}$ that is much simpler
than either of those proposed in this paper.  We have in fact performed
such an experiment, with $n=3,726,773$ \texttt{ping}'s sent from a single
source, yielding $61,246$ valid responses (for a $1.64\%$ response rate),
and resulting in an estimate $\hat{N}_{ping} = 70,583,737$.  We
then performed a \texttt{traceroute} study from the same source
to the $61,216$ unique IP addresses, and calculated a `leave-one-out'
estimate on the resulting $G^*$ of $\hat{N}_{L1O}=72,296,221$.

Of course, neither of these numbers are intended to be taken too seriously
in and of themselves.  The point is that, while the estimator from
\texttt{traceroute} data is arguably less intuitive and direct
in its derivation than that from the \texttt{ping} data, for the 
particular task of estimating $N$, it nonetheless produces 
essentially the same number.  And, most importantly, while the \texttt{ping}
data would of course not be useful for estimating $M$ or
degree characteristics, for example, the use of \texttt{traceroute}
measurements, which produce an entire sampled subgraph $\G^*$, does in
principle allow for the estimation of either of these quantities.
The success of the `leave-one-out' estimator therefore demonstrates 
both the importance and the promise of a `species'-like perspective
in the estimation of Internet characteristics.

\medskip
\centerline{\bf Acknowledgements}
\smallskip
{\small 
F.V was funded in part by the ACI Syst\'emes et S\'ecurit\'e, French
Ministry of Research, as part of the MetroSec project.
A.B. and L.D. are partially supported by the EU within
the 6th Framework Programme under contract 001907 ``Dynamically
Evolving, Large Scale Information Systems'' (DELIS).
Part of this work was performed while E.K. was with the
LIAFA group at l'Universit\'e de Paris-7, with support from the CNRS.
This work was supported in part by NSF grants CCR-0325701 and DMS-0405202 
and ONR award N000140310043.}

\centerline{\bf Appendix}

We derive here the estimator $\hat{N}_{L1O}$ of equation 
(\ref{eq:NhatL1O}).  Starting from equation (\ref{eq:Nl1out}),
and substituting $\bar{N}^*_{(-)}$ and $X$ in the numerator
for $E[N^*_{(-)}]$ and $E[X]$, respectively, our task reduces to
deriving an estimator of $(n_T-E[X])^{-1}=(n_Tq)^{-1}$, 
where $q=1-(E[X]/n_T)$.

Recall that $X=\sum_j \delta_j$ is the sum of $n_T$ Bernoulli (i.e., 0 or 1)
random variables.  If the $\delta_j$ were independent and identically
distributed (i.i.d.), with $\Pr(\delta_j=1)=p$, then $X$ would be 
a binomial random variable, with parameters $n_T$ and $p$.  
In this case, the relation $qE[(n_T+1)/(n_T+1-X)]= 1 - p^{n+1}$ holds, from
which it follows that the quantity $(n_T+1)/(n_T+1-X)$ has expectation
$q^{-1}(1-p^{n+1})\approx q^{-1}$, and therefore is an approximately
unbiased estimator of $q^{-1}$.  Substitution of the quantity
$(n_T+1)/[n_T(n_T+1-X)]$ for 
$(n_Tq)^{-1}=(n-E[X])^{-1}$ in (\ref{eq:Nl1out})
then completes the derivation of (\ref{eq:NhatL1O}).

Of course, the variables $\delta_j$ are not precisely i.i.d., due
to the commonality of sources and targets underlying the
definition of the sets $\V^*_{(-j)}$.
However, the $\delta_j$ share the same marginal distribution 
(i.e., with $p=(N-E[N^*_{(-)}])/(N-n_S-n_T+1)$\, ), and it may be 
argued that they are pairwise nearly independent under the
condition $N^*_{(-j)}\approx N^*_{(-j')}\approx N^*_{(-j,-j')}$.
These two facts together suggest that a binomial approximation to the
distribution of $X$ should be quite accurate.  It remains to 
argue for the latter fact, for which it is sufficient to show that
$\Pr(\delta_j=0|\delta_{j'}=0)
 \approx \Pr(\delta_j=0)=1-p$.  By
conditioning on the sets $\V^*_{(-j)}$ and $\V^*_{(-j')}$, counting
arguments similar to those underlying the derivation of 
equation (\ref{eq:dgivenV}) yield that
\begin{eqnarray*}
\Pr(\delta_j=0|\delta_{j'}=0) & = &
  E\left[\frac{N^*_{(-j)}-1}{N-m-n+1}\,\cdot\,
                \frac{N^*_{(-j,-j')}}{N^*_{(-j')}} \,+\,
         \frac{N^*_{(-j)}}{N-m-n+1}\,\cdot\,
                \frac{N^*_{(-j')}- N^*_{(-j,-j')}}{N^*_{(-j')}}
   \right]  \\
 & \approx & E\left[ \frac{N^*_{(-j)}-1}{N-m-n+1}\,\cdot\, 1
                     \,+\, \frac{N^*_{(-j)}}{N-m-n+1}\,\cdot\, 0 \right]
 \,\approx\, 1-p\enskip .
\end{eqnarray*}

\bibliographystyle{IEEEtran.bst}
\bibliography{IEEEabrv,biblio.bib}

\newpage

\newpage

\newpage

\newpage

\newpage

\newpage

\end{document}